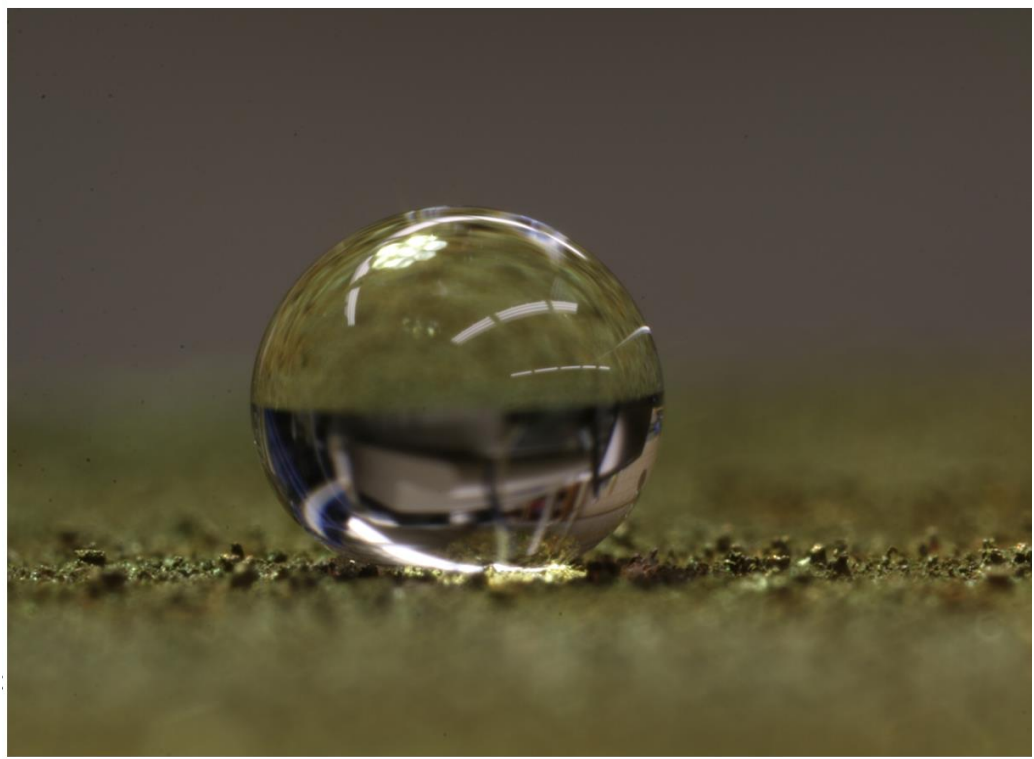

# Texture and Wettability of Metallic Lotus Leaves

C. Frankiewicz[a] and D. Attinger[a,*]

[a] Iowa State University, Black Engineering Building, Ames, IA – 50011

* Corresponding author: attinger@iastate.edu

## Abstract

Superhydrophobic surfaces with the self-cleaning behavior of lotus leaves are sought for drag reduction and phase change heat transfer applications. These superrepellent surfaces have traditionally been fabricated by random or deterministic texturing of a hydrophobic material. Recently, superrepellent surfaces have also been made from hydrophilic materials, by deterministic texturing using photolithography, without low-surface energy coating. Here, we show that hydrophilic materials can also be made superrepellent to water by chemical texturing, a stochastic rather than deterministic process. These metallic surfaces are the first analog of lotus leaves, in terms of wettability, texture and repellency. A mechanistic model is also proposed to describe the influence of multiple tiers of roughness on wettability and repellency. This demonstrated ability to make hydrophilic materials superrepellent without deterministic structuring or additional coatings opens the way to large scale and robust manufacturing of superrepellent surfaces

## Introduction

Superhydrophobic surfaces have raised great technological interest in phase change heat transfer [1], enhancing condensation [2] and boiling [3] processes, moderating the formation of ice [4] and frost [5]. A superhydrophobic surface has a low affinity to water, which corresponds to equilibrium water-air contact angles larger than ≅145°, as quantified by measurements of adhesion forces in [6]. Superhydrophobic surfaces with a large difference (hysteresis) between the advancing and receding angle cause water drops to stick on the surface [7]. On the contrary, a hysteresis smaller than ≅10° causes drops to roll off from slightly tilted superhydrophobic surfaces [8,9]. These surfaces are called superrepellent. In nature, superrepellency is observed on the leaves of several plants. It helps removing potential contaminants from the leaves (self-cleaning behavior) [10] and uncovering active pores called stomata [11]: the non-adhesion of rain and dew on the leaves [12,13] results in an improved humidification of the soil and root system, and maintains the photosynthesis and respiration processes.

It is likely that nature and technology have followed different routes to produce superhydrophobic and superrepellent surfaces. Fig. 1 relates the intrinsic wettability, depicted by the equilibrium wetting angle $\theta_E$ of water drops on a smooth slab of a given material, to the equilibrium contact angle $\theta^*$ on a textured surface of the same material. Artificial superhydrophobic surfaces are typically manufactured by texturing [14-16] a hydrophobic material using processes such as sanding, etching or lithography. Related techniques texture a hydrophilic substrate which is then coated with low surface-energy materials such as silanes [17,18], non-polar carbon [19,20] or fluorocarbons [3,21,22]. The addition of coatings can however affect thermal performance and durability [23]. These artificial surfaces are in quadrant III of Fig. 1 below the black dashed-line $\theta^*=\theta_E$ and have been called artificial lotus leaves [24,25]. Their superrepellency has also been described as *lotus effect* [26]. Fig. 1 describes the superrepellency of natural leaves, from lotus *(nelumbo nucifera)* [12], rice *(oryza sativa)* [27] and wild cabbage *(brassica oleracea)* [28]; their superrepellency has been explained by their roughness being coated with a wax layer, the cuticle [11], which was believed to be hydrophobic [12]. However wettability measurements on wax with a similar [29] composition to that of lotus leaves showed the wax to be hydrophilic rather than hydrophobic, with $\theta_E$ =74+/-9°. This situation corresponds to quadrant IV rather than III in Fig. 1. Superrepellent surfaces obtained by roughening hydrophilic materials challenge the common belief in wettability engineering that *surface roughness always magnifies the intrinsic wetting properties* [30].

Herminghaus demonstrated theoretically that a metastable superrepellent state could be obtained in quadrant IV by aggressive and fractal texturing of hydrophilic materials [31]. This theoretical result was verified by producing superrepellent surfaces from hydrophilic materials with photolithography techniques, as reviewed in [32]. For instance, deterministic pillars with overhangs were fabricated on a slightly hydrophilic silicon surface [33,34]. Similar structures were fabricated on diamond [35] and silicon [36]. Also [34], a superrepellent surface was fabricated on a hydrophilic $SiO_2$ substrate by combining re-entrant photolithographic features with nano-grass. Lithographic processes result in a high-resolution deterministic texturing of the surface and are typically suited for rather small surface areas, O($cm^2$).

In the present work, we manufacture superrepellent surfaces from a hydrophilic metal (copper) using chemical reactions rather than photolithography, and without coating the surface with low surface energy materials. Copper is a material of choice for phase change heat transfer because of a thermal conductivity significantly larger (2 to 8 times) than other common metals and silicon [37], and two to three orders of magnitude larger than typical polymers. The texture and wettability of the copper surfaces were measured and found to be comparable with those of rice leaves, lotus and brassica leaves. This random, rather than deterministic, texturing process opens the way to making superrepellent surfaces from any material that can be textured aggressively enough. By its chemical nature, the process has the ability to engineer large surface areas with dimensions ranging from a few $cm^2$ to several $m^2$, with applications ranging from computer heat sinks to airplane wings. Using a mechanistic model, we also explain why these copper surfaces are the first artificial analog of lotus leaves, and why three scales of multiscale roughness are typically needed to make hydrophilic materials superrepellent.

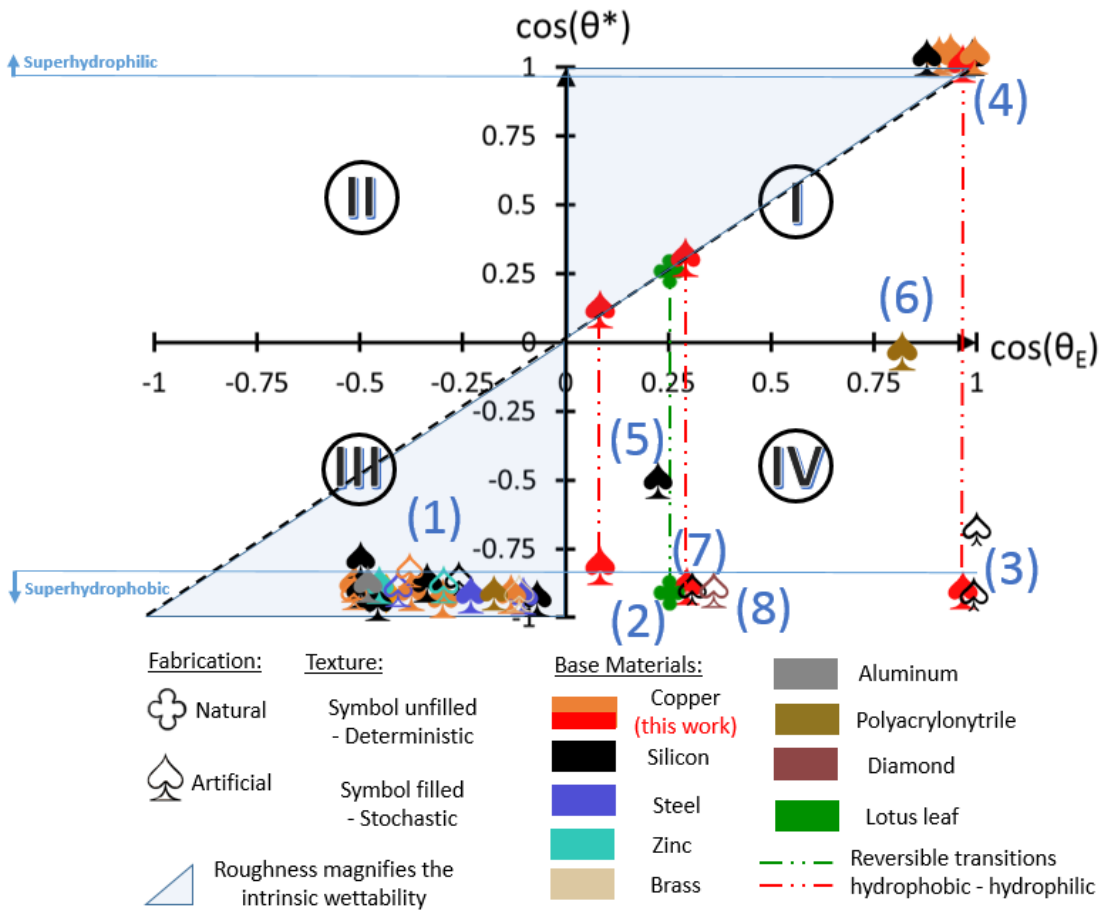

**Fig. 1** Superwettable and superhydrophobic surfaces, natural or artificial, can be classified as a function of the equilibrium wetting angles $\theta^*$ on a textured surface and the intrinsic wettability $\theta_E$ of the material to water. In quadrants I and II, hydrophilic surfaces are obtained from a bare hydrophilic or hydrophobic material, respectively. In quadrants III and IV, hydrophobic surfaces are obtained from a bare hydrophobic or hydrophilic material respectively. The line for which $\theta^*=\theta_E$ separates quadrant I and III into two areas: the blue-shaded area shows the region for which roughness magnifies the intrinsic wettability of the material. A droplet deposited on a surface with contact angle values in this blue-shaded area will be in a stable wetting state whereas outside the blue area, the wetting state will be unstable with possibilities of transitions to a stable state. The vertical dotted/dashed lines show that specific surfaces experience a transition between a metastable superepellent state, in quadrant IV, and a stable hydrophilic state in quadrant I [38]. Arabic numbers identify the following group of surfaces (1) is [17, 18, 39-49] - copper, [22, 24, 50-54] – silicon, [55-59] – steel, [60, 61] – zinc, [17, 58] – brass, [20] - aluminum, [62] - PolyAcriloNytrile (PAN); (2) [29, 36]– lotus leaf; (3) [33]– silicon; (4) [63-65] – copper, [66-68] – silicon; (5) [36] – silicon; (6) [69] – PAN; (7) [34] – Silicon. It should be noted that no smooth material exhibits a $\cos(\theta_E) < 0.6$, and also that there is no known surface in quadrant II. The work reported here focuses on superepellent surfaces in quadrant IV, using randomly textured hydrophilic materials.

## Experimental

**Surfaces fabrication**

The copper surfaces were engineered as follows. Copper samples (101 alloy, 99.99% purity, approximately 10 x 10 mm² by 3mm height) were first manually polished with a 320 grit sandpaper (average particle size of 46 μm) to remove the native oxide layer from the surface, and cleaned with isopropanol. The samples were then sonicated in a hydrochloric acid solution (5% wt. in water) for 10 to 15 minutes, and then immersed in deionized (DI) water for another 10 minutes for cleaning and removing particles due to polishing.

Then, three chemical processes were compared to modify the chemistry and texture of the copper samples.

1) Surface *E1* was prepared with an etchant according to the following process: 2mL hydrogen peroxide ($H_2O_2$, 50% wt. in water) was added to 15 mL hydrochloric acid (HCl, ACS reagent 37% titration) and stirred before adding the copper sample for micro-texturing. A solution of $H_2O_2$ and HCl etches copper by the reaction: Cu (s) + 2 HCl (aq) + $H_2O_2$ (aq) -> $CuCl_2$ (s) + $2H_2O$.

2) Surface *E2* was prepared with an etchant according to the following process: a mixture of 2g of iron chloride ($FeCl_3$, reagent grade 97%) and 15g of HCl etch micro textures in the copper. Iron chloride is a typical etchant used in the semiconductor industry for printed circuit boards [70]; it reacts with copper according to the reaction: Cu (s) + $FeCl_3$ (aq) -> CuCl (s) + $FeCl_2$ (aq). HCl was added to help decrease the etch rate and dissolve the CuCl precipitate.

3) Surface *EA* was fabricated by first following a similar etching process to *E2*, and then an additive oxidation process: we doubled the concentration of $FeCl_3$ (i.e. we added 4g) in 15 mL HCl to reduce the anisotropy and increase the pitch of the microstructures. Also, observing that CuO surfaces fabricated at 60°C had a higher roughness ratio than surfaces fabricated at 8°C, we set the temperature of reaction at 65°C, 3 degrees below the boiling point of ammonium hydroxide ($NH_4OH$). Cu reacts with ammonium hydroxide ($NH_4OH$, 28-30% wt. $NH_3$ basis in water) to form a dark black CuO layer at 60°C and a light blue $Cu(OH)_2$ layer [18] at 8°C.

After chemically processing the samples, the surfaces were cleaned in DI water for 15 minutes in a sonication water bath to remove potentially trapped chemicals and then the samples were dried in air (15 seconds under compressed air, and 15 minutes at least in ambient air) before carrying the contact angle measurements.

The above chemical processes were repeatedly used on at least four bare copper samples for each texturing process. Both the texture, the chemical composition and the wettability of the surface were found to be similar for each sample, showing the repeatability of the texturing method.

**Sourcing of leaves**

Leaves of rice *(Kitaake,* a Japanese cultivar of *oryza sativa)* and brassica (*brassica napus*) were sectioned directly from plants grown at Iowa State University. The leaves were then maintained in a humid closed container (container filled partially with water and leaves deposited on top of the water) to prevent drying. Wetting and SEM measurements (wetting, SEM) were carried out within 2 days of sectioning the leaf. Also, wettability experiments were carried out within 5 minutes after the leaves were taken out of the humid environment, and the SEM was operated in the environmental mode (ESEM) at a pressure of 120 Pa.

**Contact angle, wettability and roughness measurements**

To quantify the effect of the reaction time on the chemical modification of the copper surface, contact angle measurements were successively performed for all solutions previously mentioned: for each time plotted on the x-axis of Fig. S2, a different sample was prepared with the chemical solution, and then rinsed, sonicated for 15 minutes and dried for 15 minutes, before wettability was measured.

The static contact angle of water $\theta^*$ and the hysteresis angle $\Delta\theta$ (difference between the advancing $\theta_A$ and receding angle $\theta_R$, were measured on the solid surfaces and in ambient air using an in-house goniometer. The hysteresis was also measured by slowly controlling the volume of a spreading drop with a syringe pump (Ca << 1). The contact angles (static and dynamic) were measured at least 5 times at different locations on the surface of each sample, and averaged to the value displayed on Fig. 4. The resolution of the digital images was about 3 µm per pixel for hydrophobic surfaces and 10 µm per pixel for hydrophilic ones. Post-processing was carried out with the software ImageJ [71]. The accuracy of the contact angle measurement was ±2°.

Droplets spreading or impacting on a surface can result in the liquid penetrating (breaking into) the surface roughness. This drastically modifies the wetting state: the liquid (or some of it) sticks to the surface [4, 72]. Resistance to break-in was investigated by carrying out drop spreading and impact experiments (in Fig. S3). A critical value of the velocity, called hereafter *break-in velocity*, designates the minimum velocity for break-in to occur. For velocities lower than that break-in velocity, droplets were shown to fully bounce off of the surface, either as a whole drop or as a collection of smaller splashing droplets.

Surface structure was examined with scanning electron microscopy (SEM). The arithmetic roughness $R_a$ was measured with a 3D microscope (Hirox KH-8700, with 100nm optical resolution). More details are in the supplementary documentation.

# Results and discussion

**Texture, chemistry and wettability of selected natural and artificial superrepellent surfaces:** Fig. 2 shows a rich set of multiscale features on natural leaves and textured copper surfaces, as examined with scanning electron microscopy (SEM). The surface texture and chemistry are important to understand the surface wettability, and can be described as follows.

Natural leaves of rice (*RL*), brassica (*BL*) and lotus (*LL*) feature three scales (also called tiers [73]) of roughness. Their first tier (or largest, with width O(10-300 µm)) consists in arched structures separated by lower parallel lines for *RL* (resulting in darker areas on the SEM pictures in Fig. 2); in "tangled pretzels" shapes for *BL*, and in nipple-shaped pillars for *LL*. A second tier of roughness (O(1-10 µm)) consists of bumps for *RL*, of cracked wrinkles for *BL*, and of globules for *LL*. A third tier (O(0.01-1 µm)) is observed: nano-grass for *RL*, nano-pillars for *BL* and nano-tubes for *LL*.

Three artificial superhydrophobic surfaces have been fabricated chemically from initially polished copper slabs by etching (surface *E1*, *E2*) or by a successive sequence of etching and oxidation (*EA*), as detailed in the section "Materials and Methods". The numbers 1 and 2 indicate differences in temperature, chemical solution or processing time. The obtained surface chemistry is measured in Fig. S4: either Cu (*E1* and *E2*, with $\theta_E$=82° see Fig. 1) or CuO (*EA*, with $\theta_E$<20° see Fig. 1). Hydrochloric acid was used to decrease the degree of anisotropic etching (ratio of width over height of the micro-/nano-structures). Ammonium hydroxide was used to add a tier of roughness to copper surfaces *EA* and modify the surface chemistry by oxidation (formation of a thin CuO layer).

SEM images of engineered surfaces *E1*, *E2* and *EA* in Fig. 2 show a first tier of texture consisting of pillars, with width O(50µm). The second tier consists of micro-pillars in the case of *E1* and *E2*, and pyramidal octahedrons in the case of *EA*, O(5um). A third tier is also present on *EA* and consists of nano-pillars, O(500nm). SEM images do not reveal structures smaller than tier 2 on E1 and E2. Similarly to the natural leaves, each tier of roughness on the artificial surfaces is about one order of magnitude smaller than the former tier. A quantitative description of the width, height and pitch of these structures is provided in Table S1.

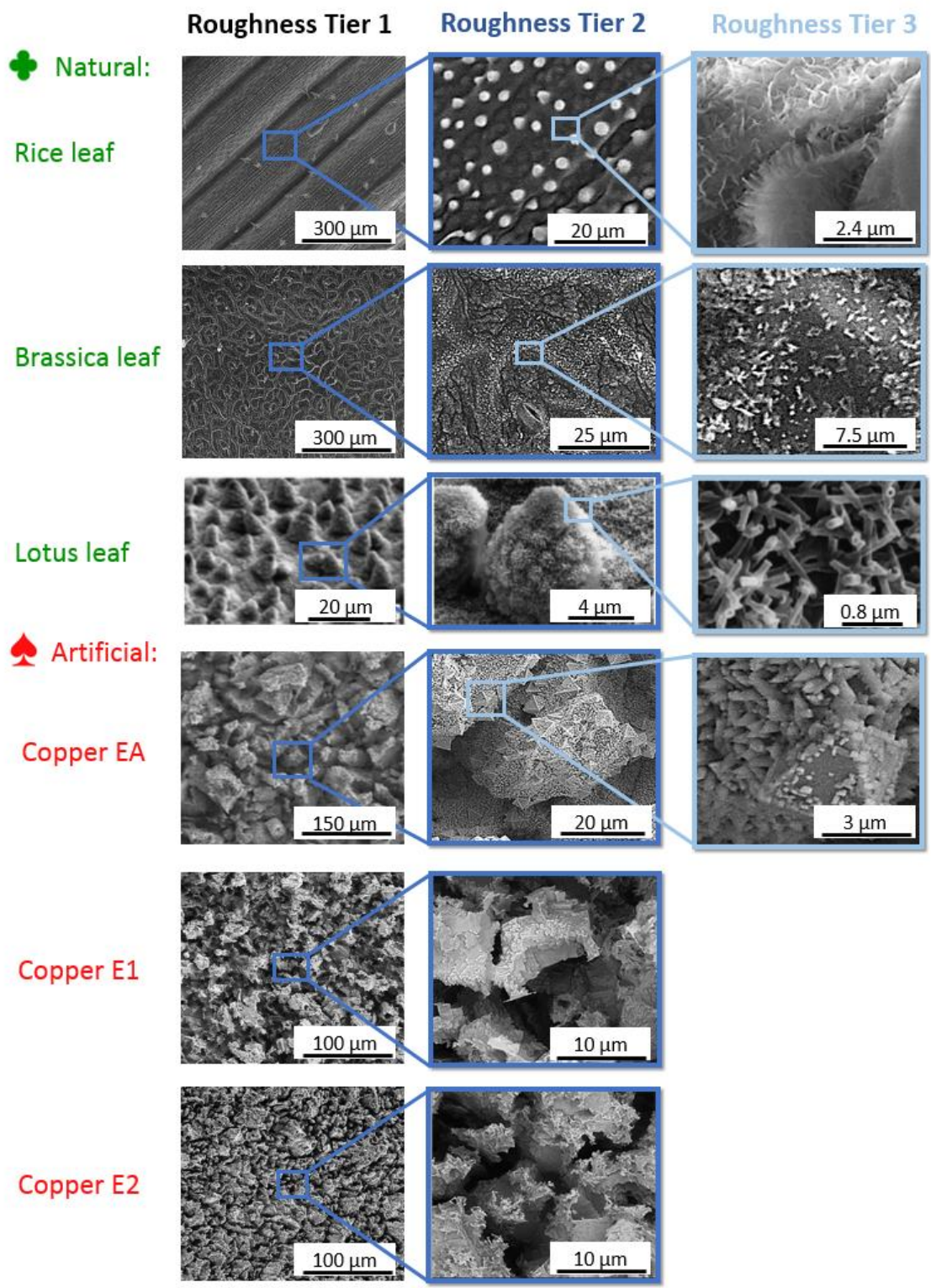

**Fig. 2** Multiscale and complex textures are observed on the superhydrophobic natural leaves (top 3 rows) and metallic surfaces described in this study. Fresh brassica and rice leaves were provided from colleagues at Iowa State University while photographs of the lotus leaf [74] are reproduced with permission.

On *EA*, *BL*, *RL*, the static contact angle was measured in this study to be $\theta^* = 160°, 164°, 161°$. The static contact angle on LL was measured as 162° in [28]. For the other artificial surfaces, the duration of the chemical processing controls $\theta^*$, measured in the ranges $82° < \theta^* < 145°$ for *E1*, and $82° < \theta^* < 143°$ for *E2*, see Fig. S2.

The hysteresis in contact angle was measured to be as high as $\Delta\theta \approx 153°$ for *E1* and $\Delta\theta \approx 146°$ for *E2*. Such large values of both $\Delta\theta$ and $\theta^*$ are typically reported on rose petals [7], with $\Delta\theta \approx 150°$ and $\theta^* = 153°$. On surface *EA* however, $\Delta\theta < 10°$. Such low values of $\Delta\theta$ on superhydrophobic surfaces have been reported for lotus and rice leaves and indicate superrepellency [27, 28].

The experimental *break-in* velocities for surfaces *E1*, *E2* are about zero m/s since the drop sticks to the surface even during slow spreading. For *EA*, the most repellent of the artificial surfaces, break-in velocity was measured as $V_0 \approx 2.1$ m/s (see supplementary information Fig. S3). The natural leaves are extremely repellent: in the case of BL and RL, even the maximum value ($V_0 \approx 5.2$ m/s) of the velocity that we could obtain experimentally (corresponding to a droplet free fall height of 1.8m) was not large enough to observe break-in.

**Analysis**

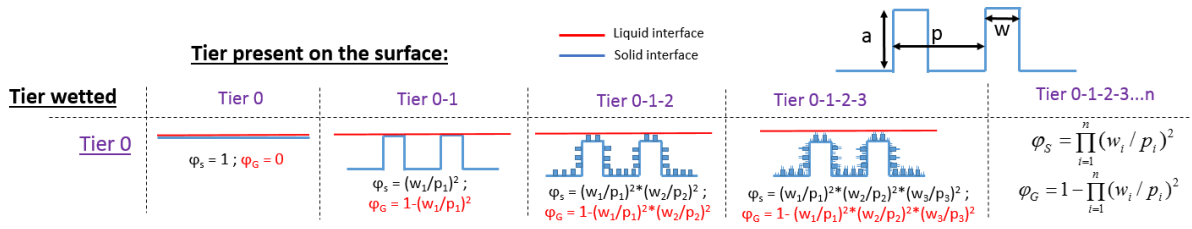

**Fig. 3** Wettability is influenced by the solid-liquid $\varphi_S$ and solid-gas $\varphi_G$ fractions at the solid-fluid interface. Here, a square wave generic model is used to estimate $\varphi_S$ and $\varphi_G$ of the complex surfaces considered in this manuscript. The model assumes a flat liquid interface. The generic texture has a number $n$ of superposed tiers (or scales) of two-dimensional square waves, defined by respective height *a*, pitch *p* and width *w*. Table S1 lists values of *a*, *p* and *w*, estimated from SEM and 3D microscopy measurements, and the corresponding $\varphi_S$ and $\varphi_G$ estimated with the square wave model in the rightmost column.

As per the above measurements and the introduction, not all superhydrophobic surfaces have the same wetting properties: surfaces *E1* and *E2* are "sticky" to water droplets similarly to rose petals, while *EA* is superrepellent like a lotus leaf. Fig. 3 and Fig. 4 provide a framework to describe the wettability of the surfaces described in this work. Their wetting state is determined [33] by three factors: the surface texture, chemistry, and the history of the exposure of the surface to the liquid. Disregarding history for the sake of simplicity for now, two theories are available to describe equilibrium wetting of a droplet on a solid surface. Wenzel [75] assumed that the liquid-solid interface follows the surface roughness, while Cassie and Baxter [76] described how air pockets, trapped in the surface roughness between the liquid and the solid, cause partial suspension of the liquid. On a hydrophilic material, Fig. 1 shows that the wetting state is either hydrophilic Wenzel or hydrophobic Cassie-Baxter, graphed in quadrant I, and IV respectively. Either situation can indeed occur on the surfaces considered in this manuscript. The hydrophilic Wenzel state is stable, and [30]

$$\cos\theta^* = r\cos\theta_E, \qquad (1)$$

with the roughness ratio *r* being the ratio between the real surface area and its normal projection. The hydrophobic Cassie-Baxter state is metastable [53,61], and

$$\cos\theta^* = \varphi_S \cos\theta_E - \varphi_G. \qquad (2)$$

Here $\varphi_S$ is the solid fraction, the ratio between the solid-liquid interface area and the horizontal projection of the solid-fluid interface [30]; while $\varphi_G$ is the gas fraction, the ratio between the air-liquid interface area and the horizontal projection of the solid-fluid interface. In the case of a flat liquid interface, equation (2) simplifies in

$$\cos\theta^* = \varphi_S(1 + \cos\theta_E) - 1. \qquad (3)$$

The complex, random geometries of Fig. 2 make estimations of *r* and $\varphi_S$ challenging, so that predicting $\theta^*$ for a given texture is difficult. To circumvent this hurdle, Fig. 3 describes a simple model to estimate the wetting angle and solid fraction of the actual surfaces, by mean of a representative generic multiscale model surface, constructed as a superposition of square wave profiles. For a surface exhibiting isotropic square waves along its two main directions, the model in Fig. 3 estimates

$$\varphi_S = \prod_{i=1}^{n} (w_i/p_i)^2, \qquad (4)$$

with symbols defined in the figure.

Fig. 4 compares the three natural and the three artificial multiscale surfaces in terms of their hydrophobicity and repellency, estimated from measurements and theory. The wettability of the surface (in terms of equilibrium angle $\theta^*$ and the hysteresis) and the break-in velocity $v_0$ are reported on the top and bottom chart, respectively. The horizontal axis indicates the number of tiers of roughness on the surface. Values of $\theta^*$ estimated with equation 3 are indicated with plain colored disks, and compared to measurements in crossed circles. Fig. 4 illustrates how $\theta^*$ and the repellency increase with the number of tiers.

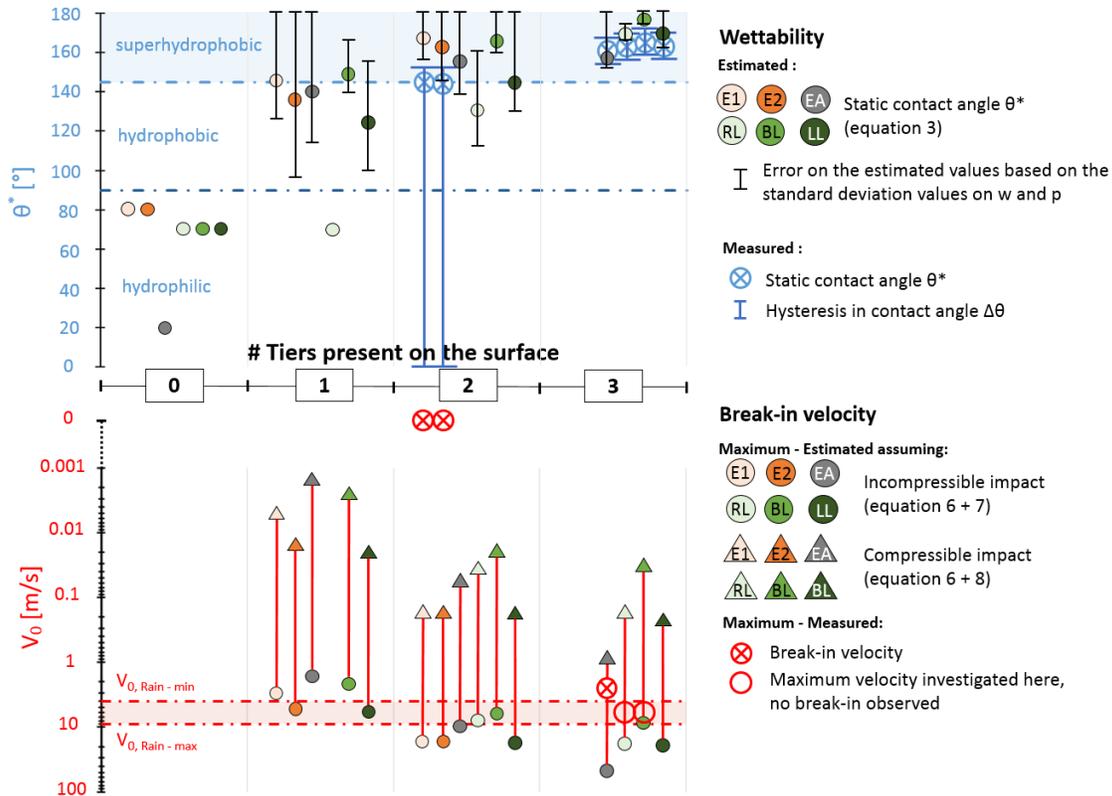

**Fig. 4: The hydrophobicity and the break-in velocity increase with the number of tiers of roughness (horizontal axis).** Top chart: the vertical axis reports the wettability as a function of the number of tiers of roughness present on the surface. Static contact angles $\theta^*$ were estimated using the generic square wave model (equation 3) together with the measured texture parameters reported in table S1. Measurements of the static contact angle and hysteresis are also reported for each surface. Bottom chart: the vertical axis reports the maximum impact velocity $V_0$ of a water droplet that the surfaces can sustain before liquid breaks into the surface texture. This break-in velocity is estimated from either equation 7 assuming that the impact is incompressible or equation 8 assuming that the impact is compressible. The break-in velocity was also measured in droplet impact experiments (reported in the supplementary information, see figure S3). Note that for the surfaces BL and RL, the break-in velocity could not be reached with the setup designed for this experiment and the values of the maximum droplet velocity reached with this setup were reported instead of the break-in velocity. The maximum and minimum values of the terminal velocity for typical raindrop sizes[77] are also reported by dashed red lines.

To understand the trends observed in Fig. 4, let us consider the ability of a multiscale roughness to maintain air pockets at the solid-liquid interface, using a mechanistic model. In the following discussion, the necessary condition for suspending a liquid on top of squared pillars will be first examined. Then, a model will be developed to determine whether the suspended liquid is in equilibrium or will submerge the roughness on the surface.

During spreading, a metastable hydrophilic Cassie-Baxter state is preferred to the stable hydrophilic Wenzel state if the surface [30] is rough enough to entrap air. This condition on the roughness ratio, for a sinusoidal profile of roughness r, is expressed for an intrinsically hydrophobic material as:

$$r = 1 + \frac{k^2 a^2}{4} \geq 1 + \frac{\tan^2(\theta_E)}{4}, \qquad (5)$$

with *k* and *Ra* being respectively the wave number and the measured roughness *Ra* (see section "Estimation of roughness and solid fraction" in the Supporting Information). For the pillar-like structures and hydrophilic values of $\theta_E$ used in this work, an analog condition[33] is met when the angle between the side of the pillars and the base surface is smaller than $\theta_E$. This explains why fully-wetting liquids bead on surfaces with reentrant structures, in a metastable Cassie-Baxter state.

The reciprocal transition, from metastable Cassie-Baxter hydrophobic to stable Wenzel hydrophilic are explained by the history of the exposure of the surface to the liquid, specifically situations where the liquid spreads on, impacts, or submerges the surface. They can be caused by air diffusion in water [29, 38]. These reversible transitions have indeed been observed for *E1*, *E2* and *EA*, reported by vertical dotted/dashed lines in Fig. 1, and for lotus leaves immersed in water [29]. More details are provided in the section "Durability" of the Supporting Information.

During drop impact, e.g. in the case where leaves are subjected to rain, there is an additional dynamic pressure driving the liquid into the surface texture, called the *break-in* [33] pressure. Break-in can either be partial, which results in an increase of the hysteresis angle and a decrease of the static contact angles for droplets in the metastable Cassie-Baxter state, or be complete (removing all the air between the liquid and the solid) and cause a transition to a stable Wenzel superhydrophilic state.

By considering the mechanical equilibrium of the triple contact line for a liquid suspended on a surface with reentrant features [33] the maximum capillary pressure resisting break-in can be estimated as

$$\Delta P_{cap} = \frac{4\sigma w_i}{p_i^2 - w_i^2}, \qquad (6)$$

with $p_i$ and $w_i$ the geometric parameters from figure 3. If an external pressure force acting on the interface, e.g. dynamic pressure from impacting droplets, is greater than the capillary pressure (i.e. $\Delta P > \Delta P_{cap}$), *break-in* occurs. If the liquid compressibility is negligible during drop impact, the *break-in* velocity is predicted as

$$V_{0,inc} = \sqrt{\frac{2\Delta P_{cap}}{\rho}}, \qquad (7)$$

with $\rho$ the liquid density. This equation was used in Fig. 4 to estimate the theoretical break-in velocity (in the case of incompressible impacts) for all surfaces. For faster impacts or rigid substrate, the liquid compressibility focuses the kinetic energy and facilitates break-in, so that [69]

$$V_{0,comp} = \frac{2\Delta P_{cap}}{\rho c}, \qquad (8)$$

with *c* the speed of sound in the liquid. This equation was used in Fig. 4 to estimate the theoretical break-in velocity (in the case of compressible impacts). On Fig. 4, the minimum and maximum values of the terminal velocity for a typical raindrop are also plotted (dashed horizontal red line, for small $D_0 \cong 1$mm, $V_0 \cong 4$ m/s and large raindrops, $D_0 \cong 7$mm, $V_0 \cong 9$ m/s [77, 78]. Estimations of the break-in velocity in Fig. 4 show how the maximum break-in velocity increases with the number of tiers present on the surface. Measurements reported in Fig. 4 confirm that all 3-tier surfaces prevent break-in significantly, while both 2-tier surfaces do not prevent break-in. Probably, the features of the 2-tier surfaces E1 and E2 are not steep or reentrant enough to suspend the interface. For surface *EA*, the measured break-in velocity is better estimated by eq. 8, which is valid for compressible impacts, while for rice and brassica leaves, the measured break-in velocity is better predicted by eq. 7, which is valid for incompressible impact. The elastic deformation of leaves, at multiples scales (see e.g. supplementary documentation) probably explains that difference. See also [79] for a study on the effect of inclination and anisotropy during impact of water drops on rice leaves. In short, surfaces are repellent to drop impact slower than their break-in velocity, a characteristic which increases with the material compliance and the miniaturization of the roughness features.

To summarize, surfaces are "superrepellent" and cause the roll off of impacting water droplets if the following three requirements are met. First, the surface must be superhydrophobic. Based on Fig. 4, the estimated and experimental values of the contact angles are equal or superior to ≈145° for surfaces with at least two tiers of roughness: this is the case for the 2-tier copper surface *E1* and *E2*, and for the 3-tier copper surface *EA* and the 3-tier leaves *RL*, *BL* and *LL*. Second, the contact angle hysteresis during spreading of water droplets must be small (typically quantified as <10°, as mentioned in the introduction), to allow for roll off and efficient droplet removal. Our measurements of contact angle hysteresis reported in Fig. 4 show that all the 3-tier surfaces *RL*, *LL*, *BL*, *EA* have a contact angle hysteresis <10° while the two-tier surfaces *E1* and *E2* have large hysteresis. Third, as mentioned in the above paragraph, the surface must prevent break-in of drops impacting in the range of velocities the surface is designed for. The estimated and measured values of break-in velocities show that only 3-tier surfaces prevents break-in of water droplet impacting up to velocities of 2.1m/s on *EA* and 5.2 m/s on *RL*, *BL*. Both 2-tier surfaces exhibit break-in even at the negligible impact velocities associated with the deposition of a sessile drop with a pipette. Fig. 4 therefore supports the argument that at least three tiers (or scales) of roughness are necessary to make hydrophilic materials superrepellent.

## Conclusion

We describe a random texturing process to fabricate the first artificial analog of the superrepellent lotus leaf, turning a hydrophilic material such as copper into a superhydrophobic, even superrepellent surface. The chemical process is compatible with large scale (O($cm^2$-$m^2$)) manufacturing of robust superrepellent surfaces. A simple mechanistic model explains how multiple tiers of roughness increase hydrophobicity and cause repellency to liquids. The model is in agreement with measurements for copper surfaces and for the brassica, rice and lotus leaves. This study indicates that at least three tiers (or scales) of roughness are needed to make hydrophilic materials superrepellent.

## Acknowledgments

We acknowledge contributions from laboratory colleagues Sahar Andalib, for helpful manuscript review, and Dr. Zaki Jubery for explanations on photosynthesis and respiration of crops, with Matthew Gilbert at UC Davis. We thank Drs. Laura Marek and Bing Yang from Iowa State University for providing us with fresh brassica and rice leaves, respectively. Financial support from the US National Science Foundation, CEBET grant 1235867, is gratefully acknowledged by both authors. The data reported in the paper are tabulated in the supplementary materials.

**Table of content (graphic)**

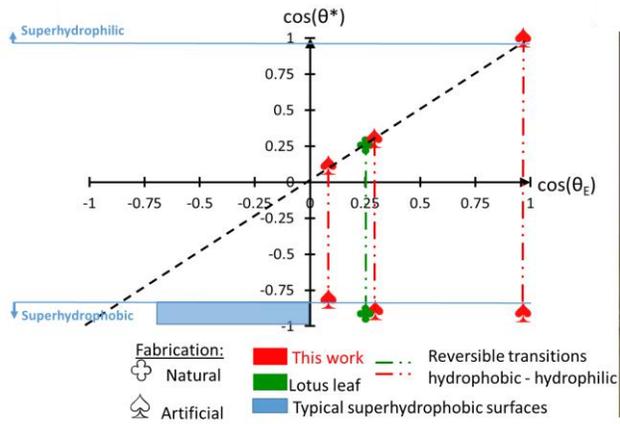 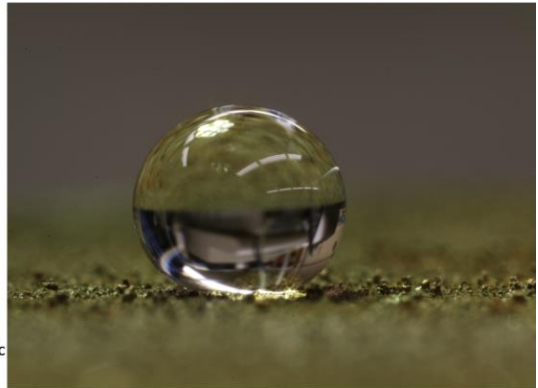

Hydrophilic materials can be made superrepellent to water by chemical texturing, a stochastic rather than deterministic process. Here, multiscale features render copper surfaces comparable to lotus leaves, in terms of wettability, texture and water repellency. The novel ability to make hydrophilic materials superrepellent without deterministic structuring opens the way to large-scale manufacturing of superrepellent surfaces.

Supplementary Information for:

# Texture and Wettability of Metallic Lotus Leaves

C. Frankiewicz[a] and D. Attinger[a,*]

[a] Iowa State University, Black Engineering Building, Ames, IA – 50011

* Corresponding author: attinger@iastate.edu

**Influence of the reaction parameters (temperature, concentration, time) on the surface texture and wettability.**

The temperature and concentration of the chemicals used for modifying the surface had a strong influence on the size and type of microstructure. Inspection of Surface E1 in Fig. S1 reveals that the lower the temperature, the higher the density of copper microstructures; and that the lower the concentration of $FeCl_3$ (Blue number, varied from 0.1g to 5g) the higher the density of copper microstructures.

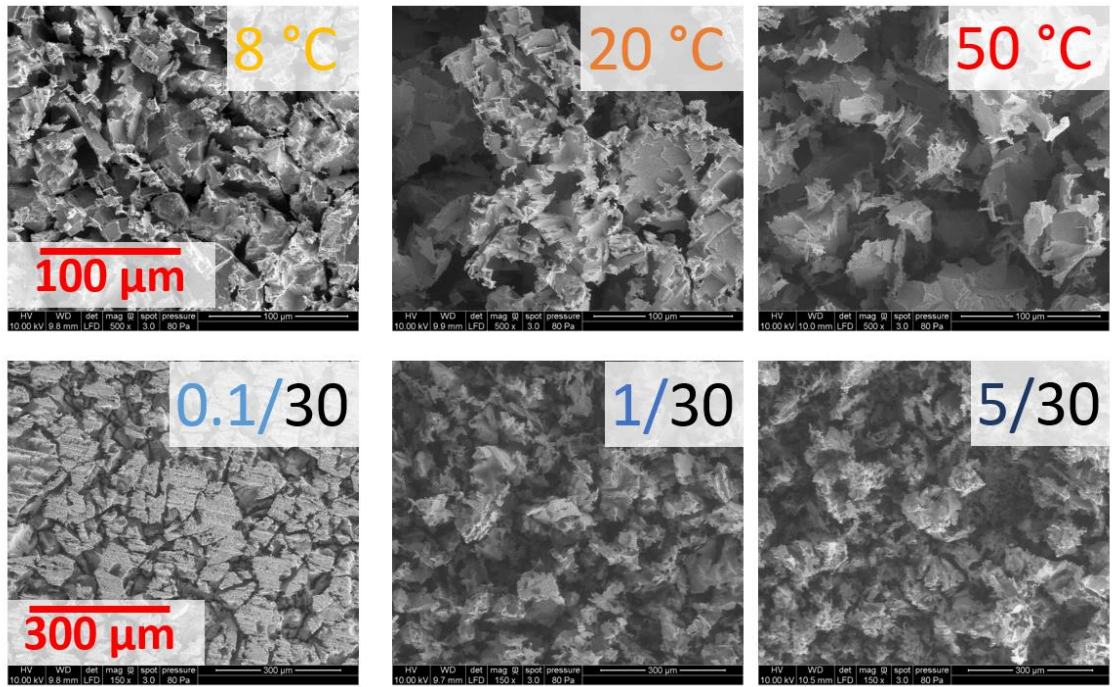

**Fig. S1: Surface E1: influence of the temperature of reaction (top row) and concentration (bottom row, in grams) of the HCl-$FeCl_3$ solution on the texture of the surface.**

The wettability and roughness are characterized in Fig. S2 for surfaces *E1* and *E2*, as a function of the time in the reactive bath. The static contact angle of a water drop on the surface $\theta^*$ was measured with a goniometer, as well as the hysteresis $\Delta\theta$ (difference between the advancing $\theta_A$ and receding angle $\theta_R$). For both surfaces, the receding angle is always equal to $\approx 0°$ and is not plotted. Superhydrophobic properties were obtained with the HCl/$FeCl_3$ (*E1*) or with the HCl/$H_2O_2$ (*E2*) etchants for reaction times longer 30h. The trend for both surfaces shows that the value of the wetting angle and roughness increase with reaction time. *EA* is not mentioned here since it is obtained by not one but two reaction times. For the surface *EA*, the value of the static and dynamic contact angles are $\theta^*$=160° and $\Delta\theta$ <10° respectively, as mentioned in Fig. 4.

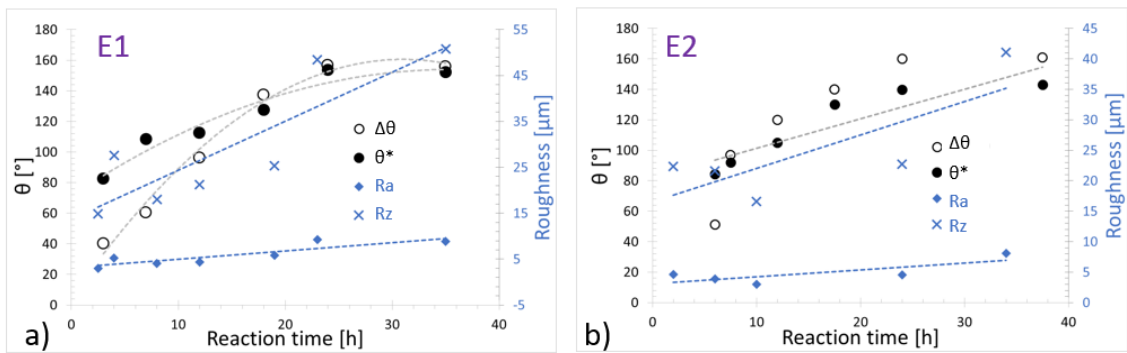

**Fig. S2: Influence of the chemical reaction time on the roughness and wettability of the fabricated surfaces.** The roughness *Ra* (arithmetic average of absolute values of the profile height deviations from the mean line) and the height *Rz* between the lowest peak and the highest valley over a 300 μm length were also plotted. The lines are only plotted as a guide for the eye.

**Estimation of roughness and solid fraction**

The solid fraction $\varphi_S$ (i.e. the ratio between the solid-liquid contact area and the projected composite area of the interface, see Fig. 3) was measured from SEM images by approximating the structures on *E1* and *E2* as cylindrical pillars (Tier 1) with cylindrical micropillars (tier 2 structures), on *EA* are cylindrical pillars (Tier 1) covered with octahedrons (Tier 2 structures) and nanopillars (Tier 3). On the natural surfaces, the first tier of roughness, structures similar to arched structures and separated by lower, darker straight lines on *RL*, "pretzels" shapes on *BL* and dome-shaped pillars on *LL* can be observed. The second tier of roughness, consists of bumps in the case of *RL*, bumps in the case of *BL* and agglomerate of tubular tubes in case of *LL*. A third tier is observed: nano-grass in the case of *RL*, nano-pillars in the case of *BL* and nano-tubes in the case of *LL*.

The values of *a*,*w* and *p* defined in Figure 3 were obtained experimentally as follows. The arithmetic roughness $R_a$ was measured 6 times at 6 different locations on the samples (one scan per location) on a ~250 µm scan length with a 3D microscope (Hirox KH-8700, with 100nm optical resolution). The average height *a* of the features for tier 1 was obtained by assuming that $a=R_a$. For tier 2 and 3, the average height *a* of the structures was measured on SEM digital images using the software ImageJ [71]. The average center to center *p* spacing between the features for each tier was measured from SEM digital images using the software Image J, assuming that the surface is wavy with wavelength $p=2\pi/k$. Finally, to obtain the width *w*, we measured the area *A* of the top of each pillar from SEM digital images using the software Image J. Then by assuming that we have cylindrical pillars, $w = 2\sqrt{A/\pi}$. The values of *p*,*a* and *w* shown in table S1 and obtained with the software ImageJ result from averaging at least 40 measurements of *p*,*a* and *w* respectively. The standard deviation on these at least 40 measurements is also reported in Table S1.

With the values of *p* and *w* shown in the Table S1, we can also estimate the solid fraction $\varphi_S$ and the gas fraction $\varphi_G$, according to the equations mentioned in Fig. 3. Results are shown in the last two columns of Table S1.

|  | Tier 1 | | | Tier 2 | | | Tier 3 | | | $\varphi_S$ | $\varphi_G$ |
|---|---|---|---|---|---|---|---|---|---|---|---|
|  | p ±Δσ [µm] | w/p ±Δσ | a/p ±Δσ | P ±Δσ [µm] | w/p ±Δσ | a/p ±Δσ | P ±Δσ [µm] | w/p ±Δσ | a/p ±Δσ | Tier 0 wetted | |
| Surface E1 | 27 ±9 | 0.35 ±0.41 | 1.87 ±0.87 | 0.81 ±0.4 | 0.36 ±0.75 | 0.62 ±1.46 | X | X | X | 0.016 | 0.984 |
| Surface E2 | 14 ±6 | 0.5 ±0.72 | 3.02 ±1.78 | 0.81 ±0.4 | 0.36 ±0.75 | 0.62 ±1.46 | X | X | X | 0.032 | 0.968 |
| Surface EA | 75 ±25 | 0.32 ±0.39 | 0.67 ±0.88 | 5.88 ±2.07 | 0.63 ±0.43 | 0.63 ±0.43 | 0.41 ±0.16 | 0.68 ±0.43 | 1.13 ±0.74 | 0.018 | 0.982 |
| Surface RL | 217 ±19 | 1 | 0.39 ±0.13 | 5.22 ±1.76 | 0.58 ±0.3 | 0.58 ±0.3 | 0.45 ±0.14 | 0.20 ±0.11 | 0.20 ±0.11 | 0.013 | 0.987 |
| Surface BL | 42 ±11.5 | 0.31 ±0.17 | 0.16 ±0.08 | 8.99 ±2.95 | 0.44 ±0.38 | 0.44 ±0.38 | 2.75 ±1.06 | 0.26 ±0.24 | 0.13 ±0.12 | 0.001 | 0.999 |
| Surface LL | 13.7 ±4.7 | 0.58 ±0.33 | 0.75 ±1.85 | 1.85 ±0.43 | 0.61 ±0.31 | 0.61 ±0.31 | 0.47 ±0.26 | 0.31 ±0.22 | 1.74 ±1.37 | 0.012 | 0.988 |

**Table S1: Values of the pitch *p*, width *w* estimated experimentally by post-processing of the SEM images in ImageJ.** Values of height *a* are taken as the roughness $R_a$ measured with a 3D microscope, with height resolution below 1 µm. $\Delta\sigma$ is the standard deviation of the experimental values (*p*, *w* and *a*). The values of $\varphi_S$ and $\varphi_G$ are also indicated.

**Drop impact measurements**

The ability of the surfaces to repel water droplets has been investigated by carrying out droplet impact experiments as shown in Fig. S3. A 7μL water droplet was repeatedly impacting on the copper surfaces bare, *E2* and *EA* at different heights, to quantify the highest sustainable pressure by the surface before break-in. Each drop volume is controlled by using a syringe pump that is connected with a plastic tubing (Internal diameter of 3mm) to a 30G needle pointing to the surface. For the experiments on the surfaces *RL* and *BL*, the leaf was maintained using a "helping hand" (mini plier) and the free fall height of the droplet was varied up to 1.8m (maximum height that can be attained by the setup). The droplet size has also been controlled from 7 μL to 25 μL to reach higher values of the impact velocity in the case of *RL* and *BL*. The camera used to capture this image sequence is a Redlake HG100K, shooting at 3000fps at a resolution of 800x600 pixels, and an exposure time of 40 μs. The surface *EA* was shown to repel water up to an impact velocity equal to 2.1 m/s (We≈150 and Re≈5000), which corresponds to a height of free fall of ≈40cm for the 7μL droplet while *E1*, *E2* and the bare copper surfaces were not shown to repel any water droplet. The surface *BL* and the surface *RL* were shown to repel a 25 μL water droplet (diameter equal to 4.8mm) even for the highest height investigated here (1.8m), which corresponds to a maximum velocity equal to 5.2 m/s (We≈1700 and Re≈23000). The compliance of the leaf and the elasticity of the microstructure may help for superrepellency.

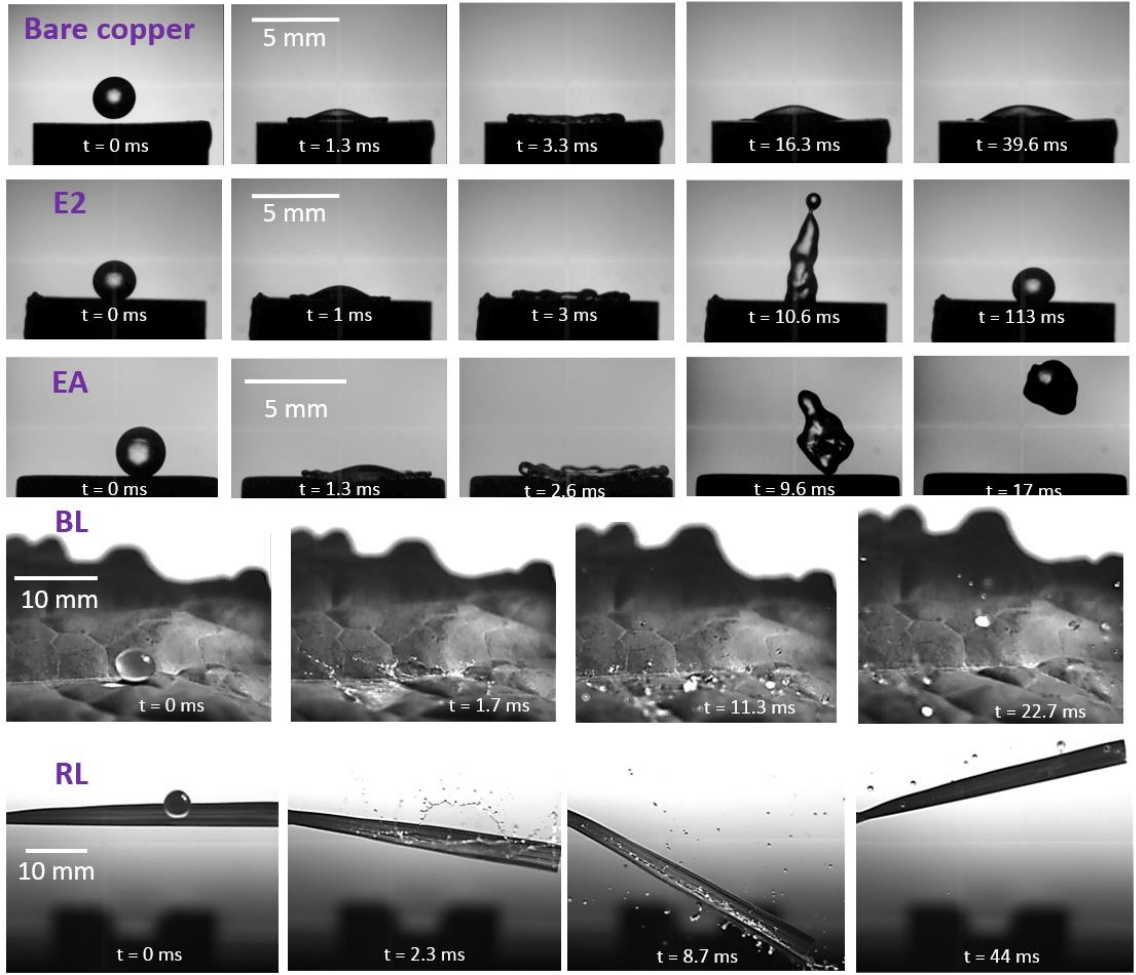

**Fig. S3: Sequence showing the comparison between the impact of a 7 μL water droplet on the bare copper surfaces, on the two engineered surfaces E2 (high hysteresis Δθ) and EA (low Δθ) falling from a height of 15 cm, and on the two leaves BL and RL for a 25 μL falling from 180cm.** During this experiment, the Weber number was We≅100 and the Reynolds number was Re≅1250. Indeed, the surface *EA* is shown to repel water up to We≅150 and Re≅5000 that corresponds to a height of free fall for the droplet of approximately 40 cm. The surface *BL* and the surface RL were shown to repel a 25 μL water droplet (diameter equal to 4.8mm) even for the highest height investigated here (1.8m), which corresponds to We≈1700 and Re≈23000.

**EDS measurements**

To confirm the chemical composition of the surfaces *E1*, *E2* and *EA* fabricated here, an Energy-dispersive X-ray spectroscopy (EDS) analysis has been carried out a few hours after fabrication and 30 days after fabrication (See Fig. S4). Surfaces *E1* and *E2* were shown to have a contact angle hysteresis that decreases with degree of oxidation. In fact, after 30 days, the contact angle hysteresis decreases from $\Delta\theta \approx 150°$ obtained after fabrication to a much lower value $\Delta\theta \approx 55°$. As shown by the EDS measurements in Fig. S4, these oxidized surfaces have a slight increase of oxygen in their chemical content, showing that an oxide has been forming on the surface (the ratio Cu/O=4/1). Indeed, with CuO surfaces (ratio Cu/O=1/1) the hysteresis is even smaller $\Delta\theta < 10°$. We can therefore suggest that the more oxidized the surface, the smaller the hysteresis. 30 days after fabrication, No change in wettability has been measured on surface EA.

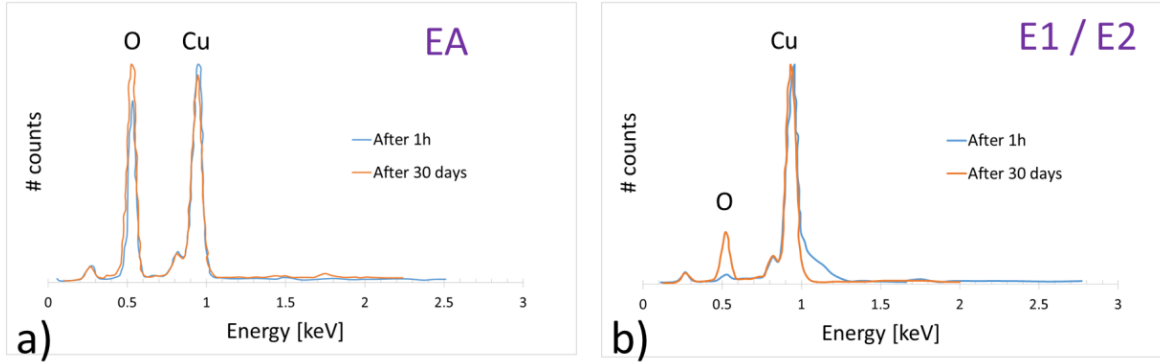

**Fig. S4: Chemical composition analysis of the fabricated samples.** a) EDS measurements of surface *EA* show that the main composition of the surface is copper oxide CuO, even after 30 days. b) EDS of surface *E2* confirm the chemical nature of Cu surfaces and its stability over 30 days. The EDS analysis of *E1* was also performed and revealed a similar surface composition (compared to *E2*) and is not shown here.

**Durability tests**

EDS measurements in Fig. S4 show that the surfaces prepared in this work are robust in terms of chemistry. Fig. S5 shows that the contact angle on surfaces *E1* and *E2* is either constant or slightly increasing, over 30 days of exposure to air. Contact angle and hysteresis were measured to remain constant for at least 30 days of exposure to air; this shows the durability of the superrepellency of surface EA.

The effect of high temperature on durability of the surface was also quantified, given that superhydrophobic surfaces are of great technological interest in boiling heat transfer[3]. Fabricated samples *E1*, *E2*, *EA* were subjected to two temperature resistance tests: (a) immersion in boiling water on a hot plate set at 150°C for one hour, and (b) heating in ambient air on a hot plate at 250°C for 10 minutes. The color, visual aspect and wetting angle values of the samples did not change after these tests.

Samples *E1*, *E2*, *EA* and one bare copper sample were also packaged for typical pool boiling experiments, as follows. A surface mount resistor was soldered on the back of each samples, with connection to a power supply (Agilent, N5750A, 750W DC). The sample was epoxied onto a Teflon casing, providing thermal insulation of its sides and back. The package is then is immersed in degassed Type II deionized water. Electrical power is then supplied to the heater to maintain nucleate boiling on the sample for at least 20 minutes, before it is set to zero W/cm$^2$ to start the pool boiling measurement. Then, the heat flux is increased by steps of $\approx$5W and maintained constant for 10 minutes before each data point is recorded. Thereafter, the heat flux is again increased at the same rate to ensure the repeatability of the experiment and obtain the value of the critical heat flux. Hysteresis between the upward and downward boiling curve was found to be negligible on all experiments. The typical duration of a boiling curve measurement was 8 hours.

Contact angles (static and dynamic) were measured on *E1*, *E2* and *EA* before and after pool boiling experiments. No significant change of wettability was observed if critical heat flux had been reached during the experiment. Samples submerged in water for 24 hours at a moderate heat flux were found to be in a hydrophilic Wenzel state, similar to the report on lotus leaf [29]. Reverting to the metastable superhydrophobic Cassie-Baxter would take about 10 minutes at 120°C, or four to seven days at atmospheric conditions (Iowa autumn is typically dry). Reaching twice the critical heat flux during pool boiling experiments did not alter wettability properties of any sample surface.

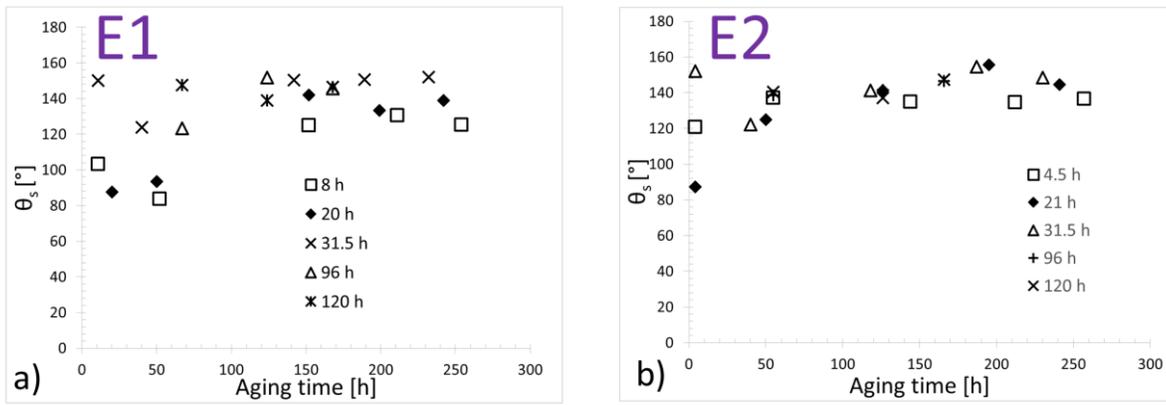

**Fig. S5: Measurement of the stability of the samples at ambient conditions.** Effect of the time after reaction (aging time on the x-axis) on the contact angle for samples *E1* and *E2*. The times mentioned as a parameter are the time of reaction of the samples. The effect of the aging time on the contact angle was also measured on *EA* without any degradation of the contact angles.